\documentclass[twocolumn]{aastex631}
\usepackage{times}
\usepackage{amsmath}
\usepackage{graphicx}
\usepackage{color}
\usepackage{array}
\usepackage{multirow}
\usepackage{amssymb}
\usepackage{color}
\usepackage{CJK}
\usepackage{float}

\newcommand{\beq}{\begin{equation}}
\newcommand{\eeq}{\end{equation}}

\newcommand{\delete}[1]{}
\newcommand{\citey}[1]{\citeauthor{#1}~\citeyear{#1}}

\begin{document}

\begin{CJK*}{UTF8}{gbsn}
\title{SN\,2025coe: A Multiple-Peaked Calcium-Strong Transient from A White-Dwarf Progenitor}
\shorttitle{SN\,2025coe}
\shortauthors{Chen et al.}

\correspondingauthor{Ning-Chen Sun}
\email{sunnc@ucas.ac.cn}

\author{Chun Chen}
\affiliation{School of Physics and Astronomy, Sun Yat-sen University, Zhuhai 519082, China}
\affiliation{CSST Science Center for the Guangdong-Hong Kong-Macau Greater Bay Area, Sun Yat-sen University, Zhuhai 519082, China}
\affiliation{Dipartimento di Fisica, Universit$\grave{a}$ di Napoli “Federico II”, Compl. Univ. di Monte S. Angelo, Via Cinthia, I-80126, Napoli, Italy}

\author{Ning-Chen Sun}
\affiliation{School of Astronomy and Space Science, University of Chinese Academy of Sciences, Beijing 100049, China}
\affiliation{National Astronomical Observatories, Chinese Academy of Sciences, Beijing 100101, China}
\affiliation{Institute for Frontiers in Astronomy and Astrophysics, Beijing Normal University, Beijing, 102206, China}

\author{Qiang Xi}
\affiliation{School of Astronomy and Space Science, University of Chinese Academy of Sciences, Beijing 100049, China}
\affiliation{National Astronomical Observatories, Chinese Academy of Sciences, Beijing 100101, China}

\author{Samaporn Tinyanont}
\affiliation{National Astronomical Research Institute of Thailand, 260 Moo 4, Donkaew, Maerim, Chiang Mai, 50180, Thailand}

\author[0000-0001-5200-3973]{David Aguado}
\affiliation{Instituto de Astrof\'{\i}sica de Canarias, V\'{\i}a   L\'actea, 38205 La Laguna, Tenerife, Spain}
\affiliation{Universidad de La Laguna, Departamento de Astrof\'{\i}sica,  38206 La Laguna, Tenerife, Spain}

\author{Ismael P\'erez-Fournon}
\affiliation{Instituto de Astrof\'{\i}sica de Canarias, V\'{\i}a   L\'actea, 38205 La Laguna, Tenerife, Spain}
\affiliation{Universidad de La Laguna, Departamento de Astrof\'{\i}sica,  38206 La Laguna, Tenerife, Spain}

\author[0000-0002-5391-5568]{Fr\'ed\'erick Poidevin}
\affiliation{Instituto de Astrof\'{\i}sica de Canarias, V\'{\i}a   L\'actea, 38205 La Laguna, Tenerife, Spain}
\affiliation{Universidad de La Laguna, Departamento de Astrof\'{\i}sica,  38206 La Laguna, Tenerife, Spain}

\author{Justyn R. Maund}
\affiliation{Department of Physics, Royal Holloway, University of London, Egham, TW20 0EX, United Kingdom}

\author[0000-0002-4870-9436]{Amit Kumar}
\affiliation{Department of Physics, Royal Holloway, University of London, Egham, TW20 0EX, United Kingdom}

\author[0000-0002-8402-3722]{Junjie Jin}
\affiliation{National Astronomical Observatories, Chinese Academy of Sciences, Beijing 100101, China}

\author{Yiming Mao}
\affiliation{National Astronomical Observatories, Chinese Academy of Sciences, Beijing 100101, China}
\affiliation{School of Astronomy and Space Science, University of Chinese Academy of Sciences, Beijing 100049, China}

\author{Beichuan Wang}
\affiliation{National Astronomical Observatories, Chinese Academy of Sciences, Beijing 100101, China}
\affiliation{School of Astronomy and Space Science, University of Chinese Academy of Sciences, Beijing 100049, China}

\author{Yu Zhang}
\affiliation{National Astronomical Observatories, Chinese Academy of Sciences, Beijing 100101, China}

\author[0000-0003-0292-4832]{Zhen Guo}
\affiliation{Instituto de F{\'i}sica y Astronom{\'i}a, Universidad de Valpara{\'i}so, ave. Gran Breta{\~n}a, 1111, Casilla 5030, Valpara{\'i}so, Chile}
\affiliation{Millennium Institute of Astrophysics, Nuncio Monse{\~n}or Sotero Sanz 100, Of. 104, Providencia, Santiago, Chile}

\author{Wenxiong Li}
\affiliation{National Astronomical Observatories, Chinese Academy of Sciences, Beijing 100101, China}

\author{C{\'e}sar Rojas-Bravo}
\affiliation{School of Astronomy and Space Science, University of Chinese Academy of Sciences, Beijing 100049, China}
\affiliation{National Astronomical Observatories, Chinese Academy of Sciences, Beijing 100101, China}

\author{Rong-Feng Shen}
\affiliation{School of Physics and Astronomy, Sun Yat-sen University, Zhuhai 519082, China}
\affiliation{CSST Science Center for the Guangdong-Hong Kong-Macau Greater Bay Area, Sun Yat-sen University, Zhuhai 519082, China}

\author[0000-0002-1094-3817]{Lingzhi Wang}
\affiliation{Chinese Academy of Sciences South America Center for Astronomy (CASSACA), National Astronomical Observatories, CAS, Beijing 100101, China}
\affiliation{Departamento de Astronom{\'i}a, Universidad de Chile, Las Condes, 7591245 Santiago, Chile}

\author[0000-0002-0025-0179]{Ziyang Wang}
\affiliation{School of Astronomy and Space Science, University of Chinese Academy of Sciences, Beijing 100049, China}
\affiliation{National Astronomical Observatories, Chinese Academy of Sciences, Beijing 100101, China}

\author{Guoying Zhao}
\affiliation{School of Physics and Astronomy, Sun Yat-sen University, Zhuhai 519082, China}
\affiliation{CSST Science Center for the Guangdong-Hong Kong-Macau Greater Bay Area, Sun Yat-sen University, Zhuhai 519082, China}

\author[0000-0001-6637-6973]{Jie Zheng}
\affiliation{National Astronomical Observatories, Chinese Academy of Sciences, Beijing 100101, China}

\author{Yinan Zhu}
\affiliation{National Astronomical Observatories, Chinese Academy of Sciences, Beijing 100101, China}

\author{David L\'opez Fern\'andez-Nespral}
\affiliation{Instituto de Astrof\'{\i}sica de Canarias, V\'{\i}a   L\'actea, 38205 La Laguna, Tenerife, Spain}
\affiliation{Universidad de La Laguna, Departamento de Astrof\'{\i}sica,  38206 La Laguna, Tenerife, Spain}

\author[0000-0003-4603-1884]{Alicia L\'opez-Oramas}
\affiliation{Instituto de Astrof\'{\i}sica de Canarias, V\'{\i}a   L\'actea, 38205 La Laguna, Tenerife, Spain}
\affiliation{Universidad de La Laguna, Departamento de Astrof\'{\i}sica,  38206 La Laguna, Tenerife, Spain}

\author{Zexi Niu}
\affiliation{School of Astronomy and Space Science, University of Chinese Academy of Sciences, Beijing 100049, China}
\affiliation{National Astronomical Observatories, Chinese Academy of Sciences, Beijing 100101, China}

\author{Yanan Wang}
\affiliation{National Astronomical Observatories, Chinese Academy of Sciences, Beijing 100101, China}



\author{Klaas Wiersema}
\affiliation{Centre for Astrophysics Research, University of Hertfordshire, Hatfield, AL10 9AB, UK}

\author{Jifeng Liu}
\affiliation{National Astronomical Observatories, Chinese Academy of Sciences, Beijing 100101, China}
\affiliation{School of Astronomy and Space Science, University of Chinese Academy of Sciences, Beijing 100049, China}
\affiliation{Institute for Frontiers in Astronomy and Astrophysics, Beijing Normal University, Beijing, 102206, China}



\begin{abstract}
SN 2025coe is a calcium-strong transient located at an extremely large projected offset $\sim$39.3 kpc from the center of its host, the nearby early-type galaxy NGC 3277 at a distance of $\sim$25.5 Mpc. In this paper, we present multi-band photometric and spectroscopic observations spanning $\sim$100 days post-discovery. Its multi-band light curves display {multiple} distinct peaks: (1) an initial peak at $t \approx 1.6$ days attributed to shock cooling emission, (2) a secondary peak of $M_{R, \, peak} \approx$  $-$15.8 mag at $t \approx 10.2$ days powered by radioactive decay, and (3) a {possible} late-time bump at $t \approx 42.8$ days likely caused by ejecta-circumstellar material/clump interaction. Spectral evolution of SN 2025coe reveals a fast transition to the nebular phase within 2 months, where it exhibits an exceptionally high [Ca II]/[O I] ratio larger than 6. Modeling of the bolometric light curve suggests an ejecta mass of $M_{\rm ej} = 0.29^{+0.14}_{-0.15} \, M_{\odot}$, a $^{56}$Ni mass of $M_{\rm ^{56}Ni} = 2.4^{+0.06}_{-0.05} \times 10^{-2} M_{\odot}$, and a progenitor envelope with mass $M_e = 1.4^{+6.9}_{-1.2} \times 10^{-3} \, M_{\odot}$  and radius $R_e = 13.5^{+64.1}_{-11.1} \, R_{\odot}$. The tidal disruption of a hybrid HeCO white dwarf (WD) by a low-mass CO WD provides a natural explanation for the low ejecta mass, the small fraction of $^{56}$Ni, and the presence of an extended, low-mass envelope.
\end{abstract}

\section{Introduction} \label{Introduction}


Over the past two decades, the advent of time-domain astronomy has led to the discovery of an increasing number of transients. Among these, Calcium-Strong Transients (CaSTs) represent a distinct class of supernovae (SNe) with low luminosity ($-14 < M_{R, \, peak} < -16.5$), rapid evolution (post-peak decline rate $\sim 0.2 $ mag/days in the $R$ band), unusually fast transition from photospheric to nebular phases ($\sim$ a few tens of  days), and diverse light-curve morphologies with single-, double-, or even multiple-peaked profiles \citep[]{De2018, Jacobson-Gal2020, Jacobson-Gal2022, Ertini2023}. CaSTs are also known as Ca-rich SNe or Ca-rich gap transients \citep[]{Perets2010,Kasliwal2012,Lunnan2017,Alsabti2017}.

A defining feature of CaSTs is their exceptionally high nebular-phase [Ca II]/[O I] flux ratios ($>$ 2), first identified by \cite{Filippenko2003, Perets2010} and subsequently confirmed by extensive studies \citep{Kasliwal2012,Lunnan2017,Alsabti2017,De2020}. {\cite{Mulchaey2014} argue that this class of transients contribute a very significant fraction of Calcium in the intracluster medium.} However, recent work has demonstrated that the strong [Ca II] emission lines do not necessarily require anomalously high calcium abundances in the ejecta. \cite{Polin2021} shows that even modest calcium mass fractions ($\sim$1\%) can produce the observed line strengths, which explains the growing preference for the name ``CaSTs", instead of Ca-rich transients, as it reflects their spectroscopic rather than nucleosynthetic distinction \citep[]{Shen2019}.  

Notably, nearly half of all CaSTs occur in early-type galaxies (E or S0 galaxies) {\citep{Alsabti2017, Scherbak2025}}, with a significant fraction exhibiting large projected offsets ($>$ 20 kpc) from their host galactic centers \citep{Alsabti2017}. The remote locations of CaSTs can potentially be explained through two channels. One scenario suggests in-situ formation of progenitor systems within the galactic halo \citep[]{Yuan2013}, though deep imaging surveys have ruled out associations with globular clusters or massive star populations at these explosion sites \citep[]{Lyman2014}. Alternatively, these systems may have been dynamically ejected from their host galaxies' central regions through gravitational interactions with supermassive BHs or supernova kicks \citep[]{Lyman2014, Foley2015}, requiring approximately 100 Myr to reach their current positions at typical ejection velocities of several hundred $\rm km \, s^{-1}$.

However, a non-negligible fraction of CaSTs also exhibit small projected offsets ($<$ 5 kpc) from their host galactic centers. Several events, such as iPTF 16hgs \citep[]{De2018} and SN 2019ehk \citep[]{Jacobson-Gal2020}, have even been found to coincide with star-forming regions of their host galaxies. {Note that there is still controversy regarding the classification of SN 2019ehk as a CaST, given that it is spectroscopically more similar to a Type IIb supernova \citep[]{De2021}.} Nevertheless, it remains unclear whether the three-dimensional physical distances of these events indeed exceed 5 kpc, as current observational evidence remains inadequate to support robust physical interpretations.

The progenitor systems of CaSTs remain unknown. The population of CaSTs with large projected offsets in early-type galaxies indicates that these events likely originate from white dwarf (WD)-related progenitors. Proposed mechanisms include: (1) the tidal disruption of a WD by a neutron star (NS) or an intermediate-mass black hole (IMBH) \citep[]{Sell2015}, (2) the explosion of a sufficiently low-mass WD \citep[]{Sim2012}, (3) the accretion-induced collapse (AIC) of a WD \citep[]{Darbha2010}, (4) the well-studied `.Ia model' involving helium shell detonation on the WD surface where the shock propagation fails to ignite the core \citep{Bildsten2007, Shen2010, Waldman2011, Woosley2011, Sim2012, Alsabti2017}, (5) the double-WD merger model \citep[]{Bobrick2017, Perets2019, Zenati2019}. Another group of CaSTs located near galactic centers or within star-forming regions are considered to result from massive stellar explosions with completely stripped envelopes \citep[]{Nakaoka2021, De2021}. Therefore, resolving the origin of CaSTs remains an outstanding problem in time-domain astronomy.

\begin{figure}[!t]
\centering
\includegraphics[scale = 0.5]{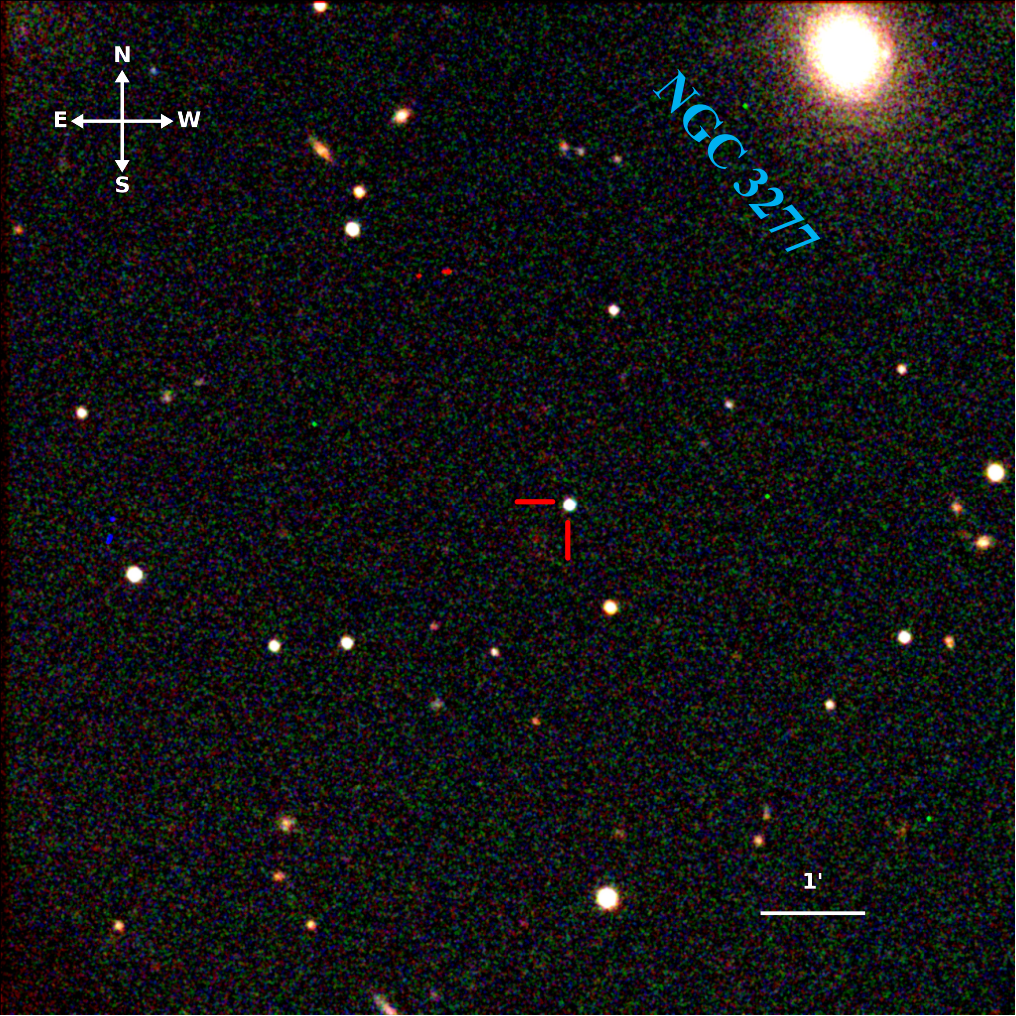}
\caption{SN 2025coe and its host galaxy NGC 3277. The image is a composite of BV$R$-band observations obtained by TRT at $t = 1.7$ days after the discovery. Red cross marks the location of SN 2025coe.}
\label{host}
\end{figure}

SN 2025coe is a newly-discovered CaST in the nearby universe, distinguished by its multi-peaked light curve. It was first discovered by \citet{Itagaki2025} on 2025 February 24 (MJD 60730.63) with a magnitude of 17.4 in the clear band at J2000 coordinates $\alpha = 10^h33^m07.95^s, \delta = +28^\circ\,26'\,13.10''$ (see Figure \ref{host}). The last non-detection before discovery was recorded on MJD 60729.5, i.e. $\sim$1\,days before discovery, in the ATLAS $o$-band. The host galaxy of SN 2025coe is NGC 3277, classified as an SAa/SAb-type galaxy with a redshift of z = 0.00472, corresponding to a luminosity distance of $D_L \approx$ 25.5 Mpc \citep[]{Mould2000}. SN 2025coe is located at a projected distance of $\sim 317.7''$ ($\sim$ 39.3 kpc) from the nucleus of its host.

We initiated multi-band photometric and spectroscopic monitoring of SN 2025coe beginning approximately 1.5 days after its discovery. The combination of its close distance, early discovery with rapid follow-up observations, and remarkable features including a multi-peaked light curve and substantial offset from the host galactic center provides an exceptional opportunity to advance our understanding of both the explosion mechanisms and progenitor systems of CaSTs. In this paper, we present detailed results of our study.


Throughout the paper, all epochs are relative to discovery (MJD = 60730.63) unless otherwise specified. Given that SN 2025coe is located far from its host galaxy, we only applied Galactic extinction corrections to the photometry; following dust maps in \cite{Schlafly2011}, we adopt $E(B-V) = 0.024$ and use the standard reddening law with $R_V = 3.1$ for the Galactic extinction correction \citep[]{Fitzpatrick1999}.

The paper is organized as follows. In Section \ref{observation}, we describe our long-term multi-band photometric and spectroscopic monitoring. We provide a detailed analysis of these multi-wavelength observations in Section\ref{analysis}. In Section \ref{LC}, we construct the bolometric light curve of SN 2025coe and perform model fitting. In Section \ref{Discussion}, we discuss our investigation into the progenitor system of SN 2025coe, with conclusions summarized in Section \ref{Conclusion}.

\section{Observations} \label{observation}

\subsection{Photometry}

\begin{figure*} 
\centering
\includegraphics[scale = 0.5]{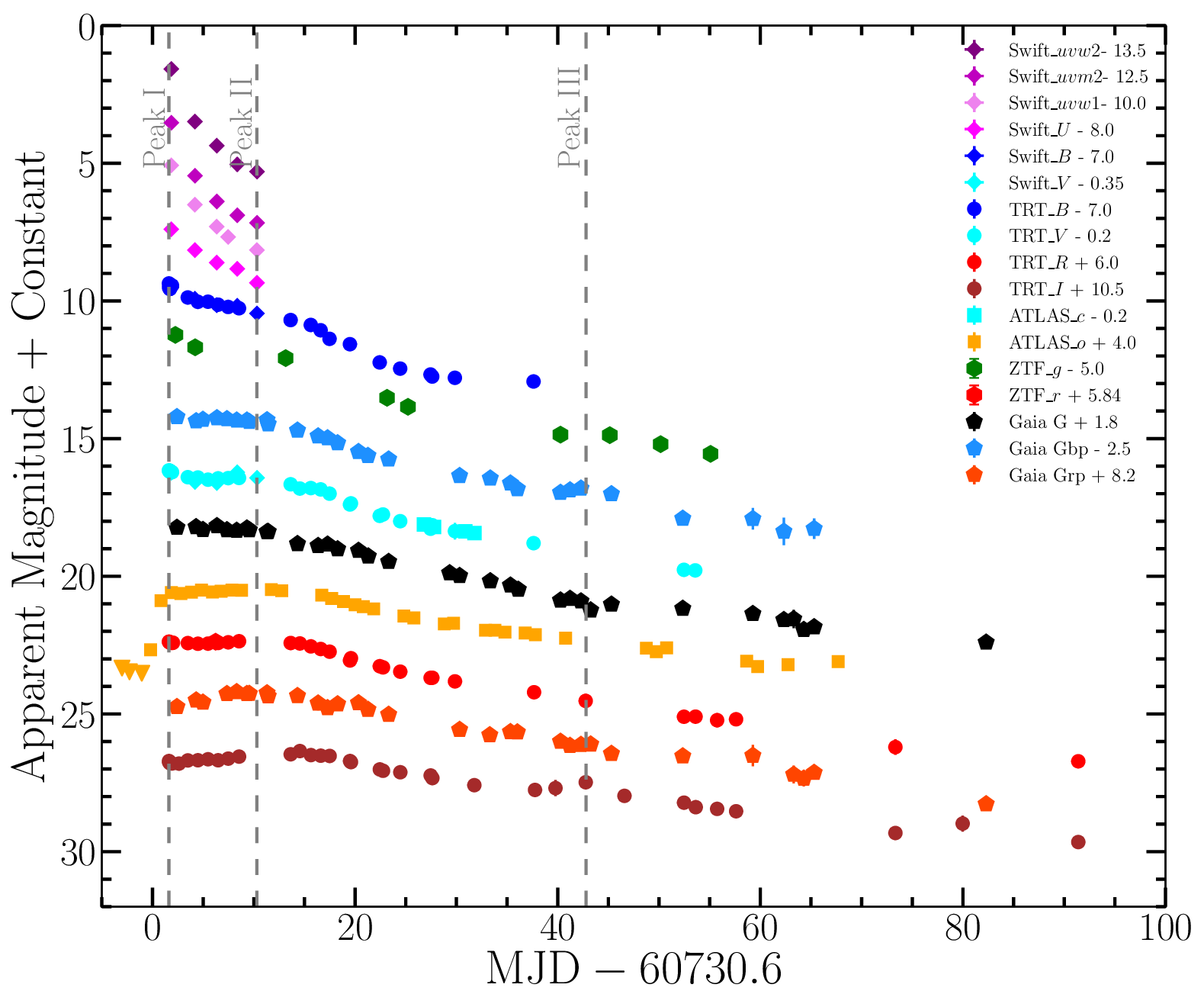}
\caption{The multi-band light curves of SN 2025coe are shown with phases relative to the first clear-band detection (MJD 60730.6). Gray dashed lines mark the epochs corresponding to the {multiple} photometric peaks. No extinction corrections are applied.}
\label{phot_sum}
\end{figure*}

We triggered Target-of-Opportunity (ToO) follow-up observations of SN2025coe with the Ultraviolet Optical Telescope (UVOT; \citey{Roming2005}) on board the Neil Gehrels Swift Observatory \citep[]{Gehrels2004} starting from 26 February 2025 until 06 March 2025 UTC ($t = 1.9 - 10.3$ days since discovery), with a cadence of 2 days using a sequence of filters including $uvw2$, $uvm2$, $uvw1$, $U$, $B$, and $V$. We conducted aperture photometry using a $3''$ aperture region with the \texttt{uvotsource} tool in \texttt{HEAsoft} v6.35 (along with the corresponding calibration files), following the standard procedures outlined in \cite{Brown2014}. Given the isolation of SN 2025coe, no background subtraction is required. Results of photometry are listed in Table \ref{tbl:phot_table_s}. All magnitudes for Swift/UVOT are given in the Vega photometric system.

\begin{deluxetable}{cccccc} 
\tablecaption{UV-Optical photometry  results of SN~2025coe. Here we only show the first 10 rows; the full table is available online. 
\label{tbl:phot_table_s}}
\tablecolumns{6}
\tablehead{
\colhead{MJD} &
\colhead{Phase\tablenotemark{a}} &
\colhead{Filter} & \colhead{Magnitude} & \colhead{Uncertainty} & \colhead{Instrument}
}
\startdata
60730.4 & -0.2 & $o$ & 18.67 & 0.05 & ATLAS \\
60731.4 & +1.0 & $o$ & 16.89 & 0.01 & ATLAS \\
60732.2 & +1.6 & $B$ & 16.37 & 0.12 & TRT \\
60732.2 & +1.6 & $V$ & 16.35 & 0.06 & TRT \\
60732.2 & +1.6 & $R$ & 16.38 & 0.07 & TRT \\
60732.2 & +1.6 & $I$ & 16.21 & 0.10 & TRT \\
60732.3 & +1.7 & $B$ & 16.56 & 0.04 & TRT \\
60732.3 & +1.7 & $V$ & 16.40 & 0.26 & TRT \\
60732.3 & +1.7 & $R$ & 16.41 & 0.07 & TRT \\
60732.3 & +1.7 & $I$ & 16.29 & 0.13 & TRT \\
\hline
\enddata
\tablenotetext{a}{Relative to discovery (MJD 60730.6)}
\end{deluxetable}




Besides, soon after the discovery of SN 2025coe, we rapidly triggered the Thai Robotic Telescope (TRT), obtaining the first data point at t = 1.6 days using a set of BVRI filters. The observations were conducted from its sites at the Sierra Remote Observatories, the Cerro Tololo Inter-American Observatory, and the Gao Mei Gu Observatory. Over the first 30 days, we use a typical cadence of 1--2 days. A lower cadence is adopted after $t \approx 30$ days and we continue the follow-up observations out to $t = 91.4$\,days after which SN2025coe falls below the detection limit. Standard data reduction was applied, including bias subtraction and flat-field correction. Point-spread-function (PSF) photometry was performed using \texttt{AutoPHOT} \citep[]{Brennan2022}. Photometric calibration was performed with the APASS catalog \citep[]{Henden2009}, converted into the Johnson-Cousins system using the method in \cite{Lupton2005}.


The RAPAS ProAm network \cite[]{Thuillot2022} also conducted intensive monitoring of SN 2025coe from 2025 February 28 to May 29. They acquired 59 high-quality photometric measurements in their specialized filter system (RAPAS A/B/C bands precisely matched to Gaia DR3's G/Gbp/Grp passbands). These data are released to the public and we retrieved their data from the Transient Name Server \footnote{\url{https://www.wis-tns.org/object/2025coe}}.


In addition, we also retrieved the {ATLAS \citep[]{Tonry2018a, Tonry2018b} and ZTF \citep[]{Bellm2019, Graham2019}} data for SN 2025coe via the forced-photometry services \footnote{\url{https://fallingstar-data.com/forcedphot/}} \footnote{\url{https://ztfweb.ipac.caltech.edu/cgi-bin/requestForcedPhotometry.cgi}}. These data points are also included in Table \ref{tbl:phot_table_s}.

\subsection{Spectroscopy}


Spectroscopic monitoring of SN 2025coe was conducted from $t = 0.9$ days to $t = 87.4$ days. Table \ref{Log_of_spectra} shows a complete log of spectroscopy, which includes our observations conducted with Xinglong 2.16-m telescope (XL-216), Liverpool Telescope (LT) and Gran Telescopio Canarias (GTC), as well as publicly available spectra from Las Cumbres Observatory (LCO) network \citep[]{Brown2013}, retrieved from the Transient Name Server.

The data reduction for the spectra from XL-216 and LT follows the standard pipeline, including bias subtraction, flat-field correction, sky background removal, 1D spectral extraction, wavelength calibration using comparison arcs, and flux calibration with standard stars. The raw data of the spectra from GTC were processed using the \texttt{PypeIt} pipeline \citep[]{Prochaska2020}. We applied photometry-based flux corrections to all the spectra to improve the flux calibration accuracy. 

\begin{figure} [!t]
\centering
\includegraphics[scale = 0.5]{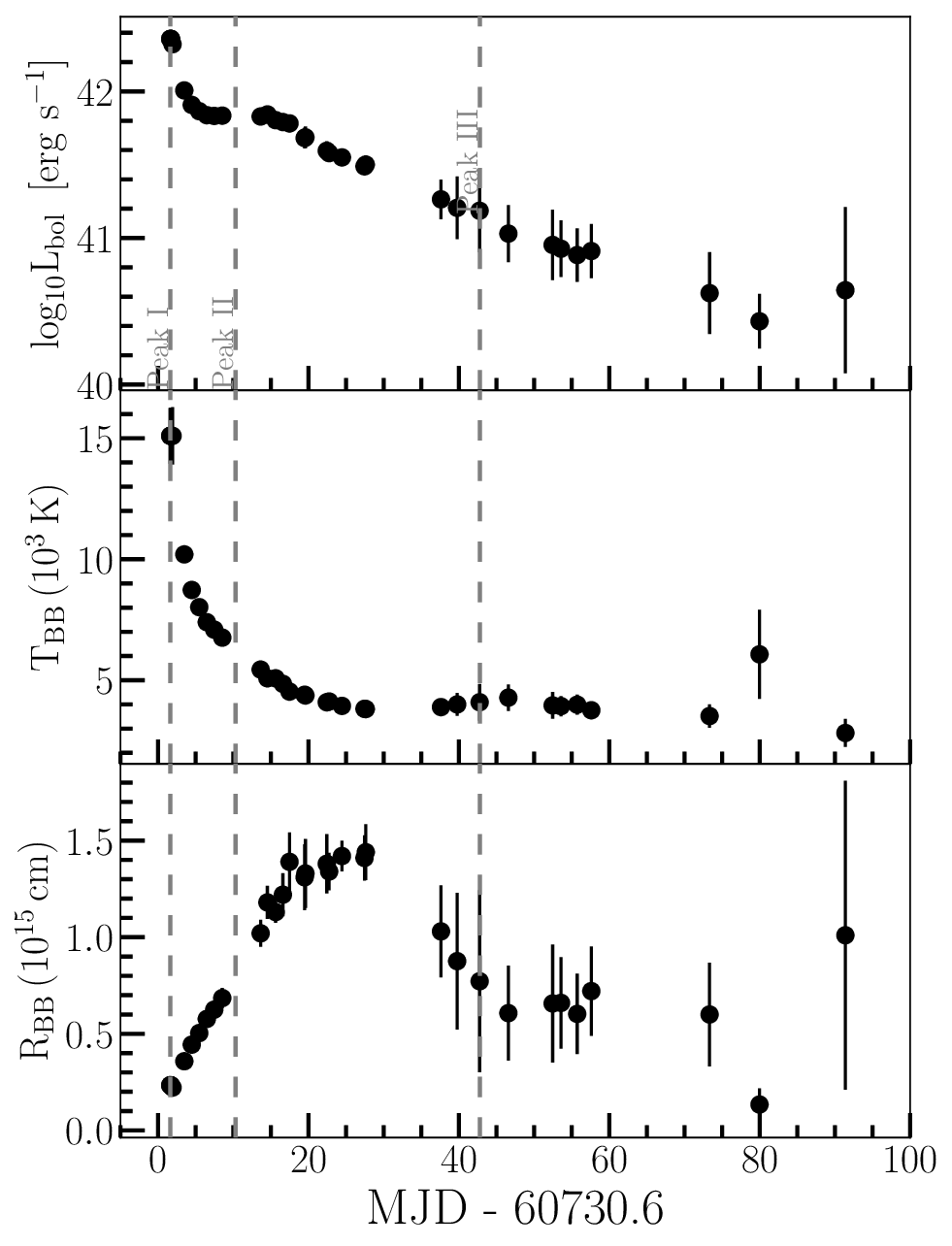}
\caption{Top panel: Bolometric light curve of SN 2025coe derived from fitting the multi-band photometric results. Middle panel: Temporal evolution of the blackbody temperature. Bottom panel: Temporal evolution of the blackbody radius. The gray dashed lines mark the epochs of the {multiple} peaks detected from the multi-band light curves.}
\label{BB_fit_res}
\end{figure}

\begin{table*} 
\centering 
\caption{Optical Spectroscopy of SN 2025coe. The phase is indicated in the discovery SN 2025coe.}
\begin{tabular}{cccccc} \hline 
\hline
MJD & Phase (d) & Filter & Grism & Spectral range ($\mathrm{\AA}$) & Telescope/Instrument \\ \hline
60731.5 & 0.9 &  & & 3200 - 5700, 5400 - 10000 & LCO/FLOYDS-N \\ 
60732.1 & 1.5 &  & & 4020 - 8100  & LT/SPART \\
60732.8 & 2.2 & 385LP  &  G4 &  3700 - 8800 & XL-216/BFOSC \\
60733.0 & 2.4 &  & & 4020 - 8100  & LT/SPART \\
60737.5 & 6.9 &  & & 3200 - 5700, 5400 - 10000 & LCO/FLOYDS-N \\
60738.8 & 8.2 & 385LP  &  G4 &  3700 - 8800 & XL-216/BFOSC \\
60740.5 & 9.9 & 385LP  &  G4 &  3700 - 8800 & XL-216/BFOSC \\
60742.5 & 11.9 & 385LP  &  G4 &  3700 - 8800 & XL-216/BFOSC \\
60751.2 & 20.6 &  & & 3200 - 5700, 5400 - 10000 & LCO/FLOYDS-N \\
60752.0 & 21.4 &  & & 4020 - 8100  & LT/SPART \\
60768.0 & 37.4 &  & & 4020 - 8100  & LT/SPART \\
60783.9 & 53.3 &  & & 4020 - 8100  & LT/SPART \\
60790.9 & 60.3 &  & & 4020 - 8100  & LT/SPART \\
60818.0 & 87.4 &  & R1000R &  5100 - 10000  & GTC/OSIRIS+ \\
\hline
\end{tabular}
\label{Log_of_spectra}
\end{table*}


\section{Analysis} \label{analysis}
\subsection{Light Curve Properties}

SN 2025coe was last non-detected in the ATLAS $o$-band at MJD 60729.5. It was subsequently detected by ATLAS on MJD 60730.4. This constrains the explosion time of SN 2025coe to within one day. Shortly afterward, the astronomer Itagaki reported it at MJD 60730.6. We obtained the first multi-band photometric data points for SN 2025coe at $t$ = 1.6 days with TRT in the $BVRI$ bands, and at $t$ = 1.8 days with Swift in the $uvw2$, $uvm2$, $uvw1$, $U$, $B$ and $V$ bands. The first-epoch data across these bands indicate that SN 2025coe was very bright in the short-wavelength regimes at this early phase, suggesting the presence of a hot radiating source.

Following the initial observations, the $uvw2$, $uvm2$, $uvw1$, U and B bands exhibit a fast decline in brightness, while the longer-wavelength bands continue to rise until $t \approx$ 10.2 days, when the $R$-band reaches its peak. Thereafter, the luminosity declines across all filters. However, at approximately $t \approx$ 42.8 days, a possible third peak emerges in the $g$/Gbp/G/Grp/$I$ bands, although this feature is not significant in other filters where the light curve sampling is sparse. In this paper, we refer to the second peak (at t $\approx$ 10.2 days) as the main peak.


\subsubsection{Bolometric Light Curve}

 We derived the evolution of the bolometric luminosity $L_{\rm bol}$, blackbody temperature $T_{\rm BB}$ and radius $R_{\rm BB}$ with the \texttt{superbol} tool \citep[]{Nicholl2018}. 
Figure \ref{BB_fit_res} shows the results, where we have marked the epochs of the {multiple} distinct peaks as mentioned in Section 3.1. 

The first peak of the bolometric light curve reaches a luminosity of approximately $2.3 \times 10^{42} \, \rm{erg \, s^{-1}}$, with a corresponding $T_{\rm BB} \simeq$ 15,000 K and a $R_{\rm BB} \simeq 2.3 \times 10^{14}$ cm, indicating an initially very hot and luminous emitting region. After the first peak, the luminosity shows a rapid decline out to $t = $ 5 days followed by a possible plateau near the main peak from $t$ = 5 to 15 days. At the same time, $T_{\rm BB}$ cools down from 10,000 K to 6,000 K, while $R_{\rm BB}$ expands from $5.0 \times 10^{14}$ cm to $1.1 \times 10^{15}$ cm.

From $t =$ 15 days, the luminosity decreases almost exponentially out to $\sim 40$ days. The possible third peak visible in the \textit{g}/Gbp/G/Grp/I-band light curves cannot be clearly seen in the bolometric light curve. This is because the multi-band monitoring of SN 2025coe has a very low cadence and covers only a limited number of filters during this late phase, resulting in large uncertainties in the derived bolometric luminosity at that epoch. Despite the large bolometric uncertainties, $T_{\rm BB}$ exhibits a slight increase at the 3rd peak while $R_{\rm BB}$ remains nearly constant at $\sim 8.0 \times 10^{14}$ cm.

\subsubsection{Comparison with other CaSTs}



\begin{figure*} 
\centering
\includegraphics[scale = 0.35]{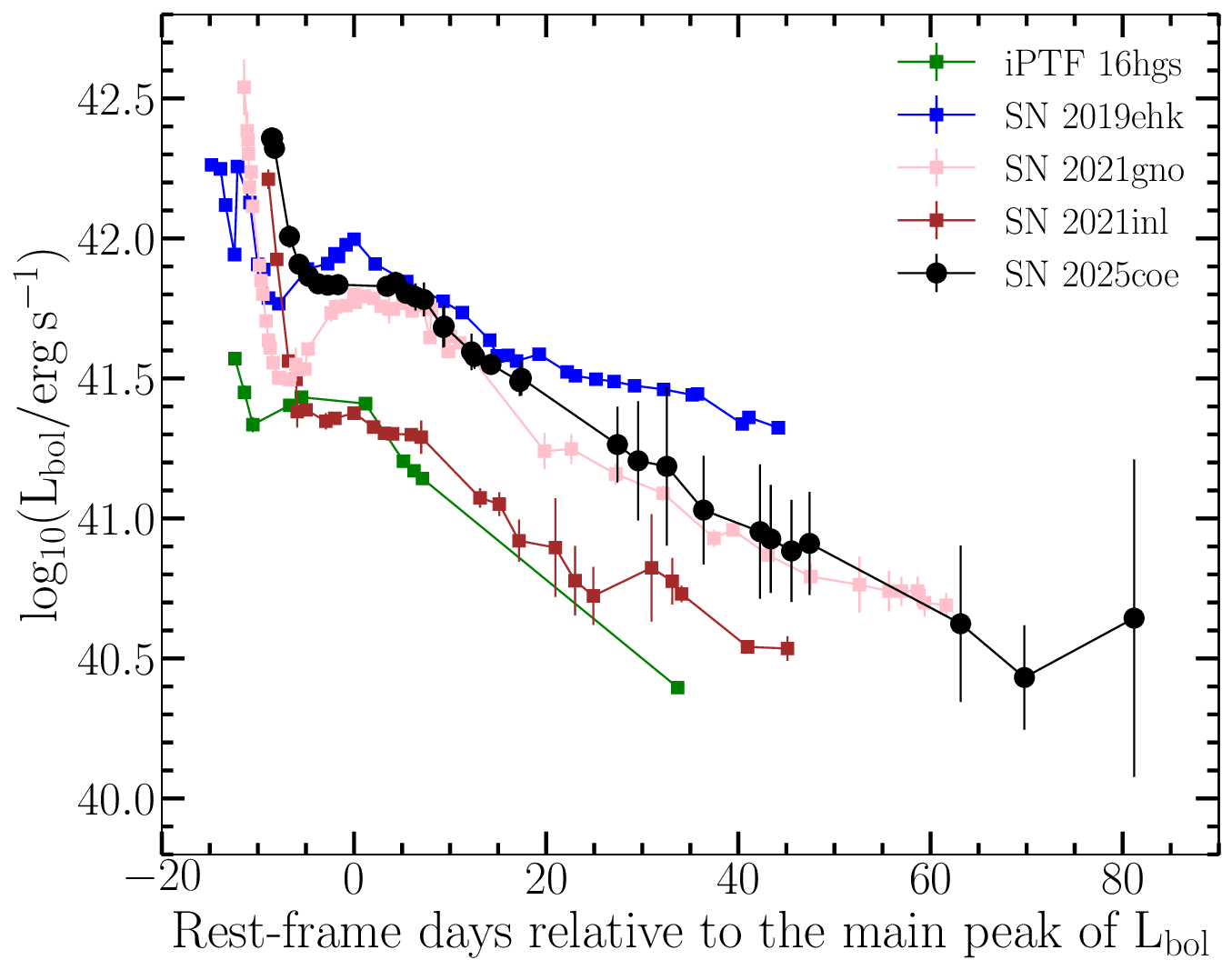}
\includegraphics[scale = 0.35]{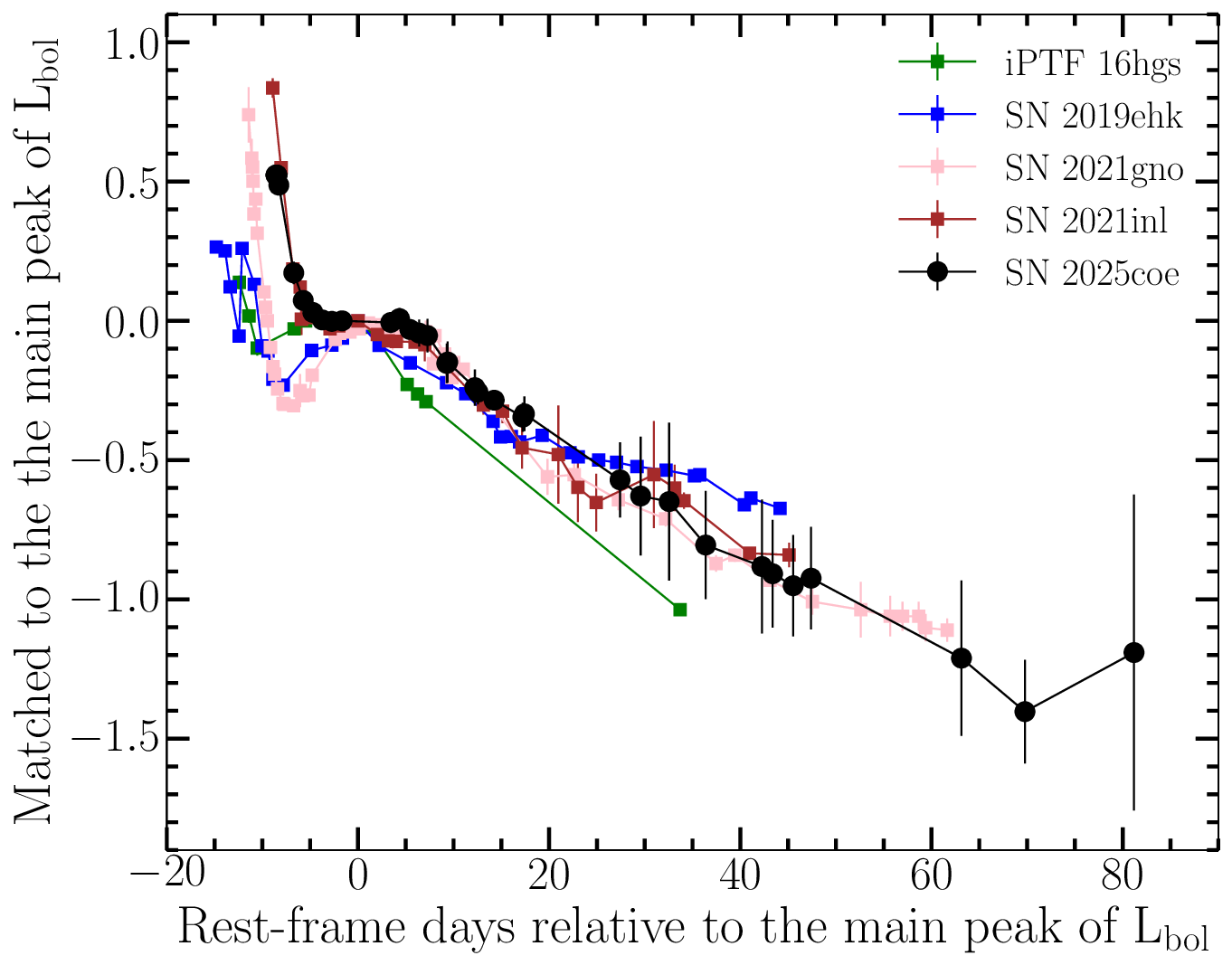}
\includegraphics[scale = 0.35]{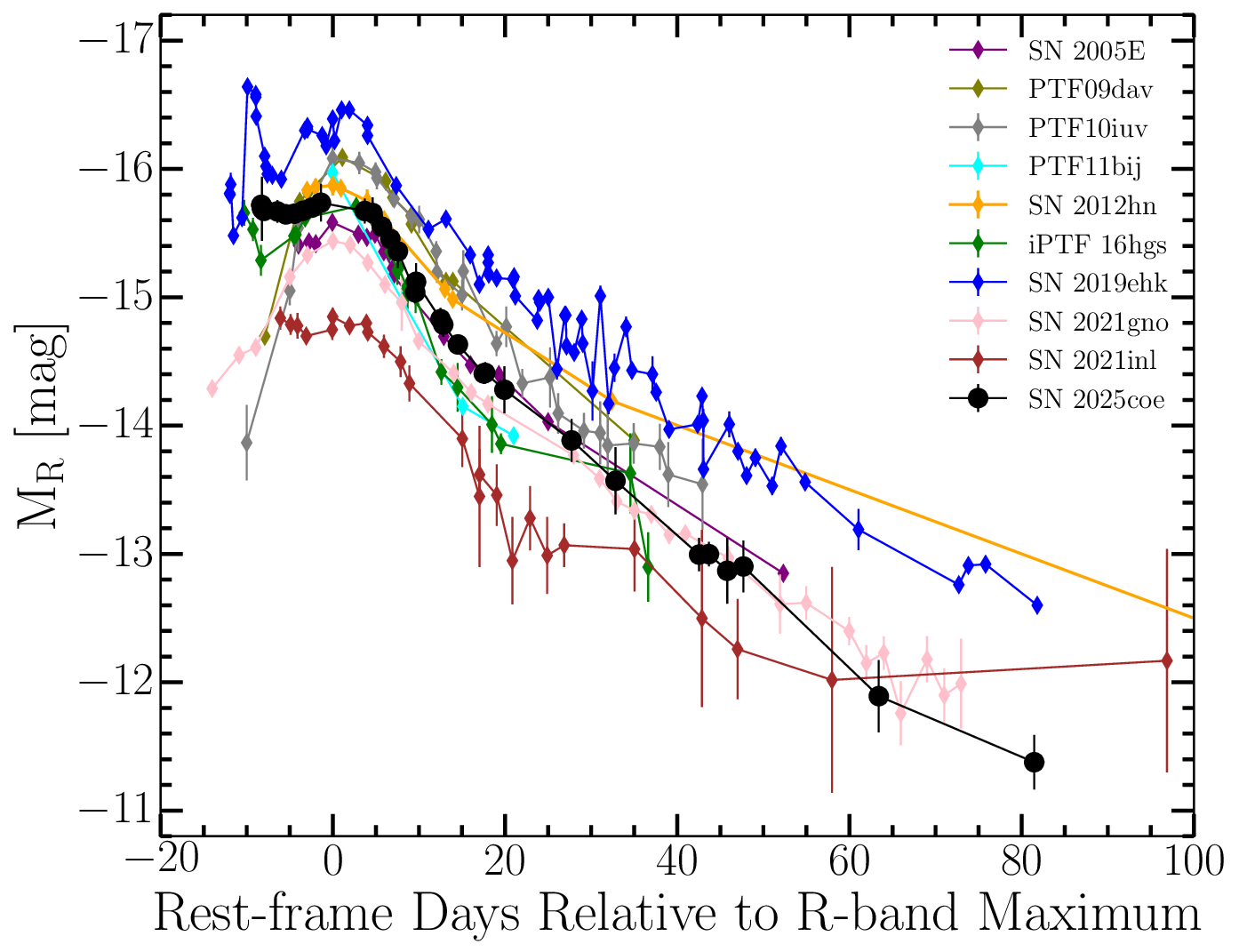}
\includegraphics[scale = 0.35]{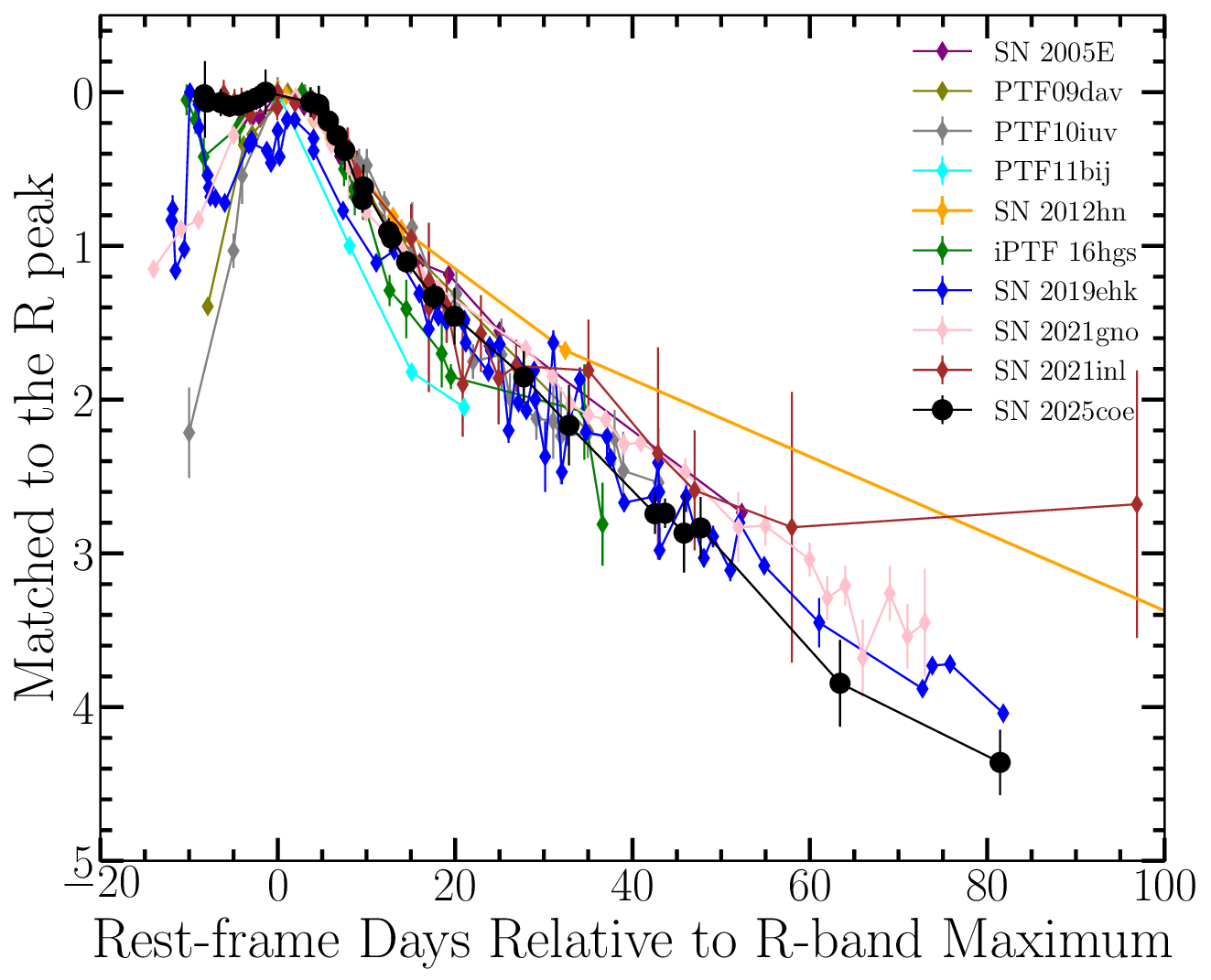}
\caption{Top left: Bolometric light curve of SN 2025coe (black dots) compared to other multi-peaked CaSTs. Top right: Scaled bolometric light curves (normalized to the main peak luminosity). Bottom left: $R$-band light curve of SN 2025coe (red dots) compared to other CaSTs (gray diamonds). Bottom Right: scaled $R$-band light curves (normalized to the peak luminosity).}
\label{Lbol_com}
\end{figure*}

We compare the bolometric and the $R$-band light curves of SN 2025coe with those of other CaSTs \citep[]{Perets2010,Kasliwal2012,De2018,Jacobson-Gal2020,Jacobson-Gal2022,Ertini2023} in Figure \ref{Lbol_com}. The bolometric luminosity of SN 2025coe is relatively high compared to other multi-peaked CaSTs, while the main peak luminosity of SN 2025coe in R band is typical compared to that of other CaSTs. By fitting the $R$-band light curve of SN 2025coe with a polynomial function, we determine that the peak luminosity of the main peak is $M_{R, \, peak} \approx -$15.8 mag. The decline rates of both the bolometric and $R$-band light curves are typical among CaSTs.

\subsection{Spectroscopic Properties}

\begin{figure*}[t!] 
\centering
\includegraphics[scale = 0.5]{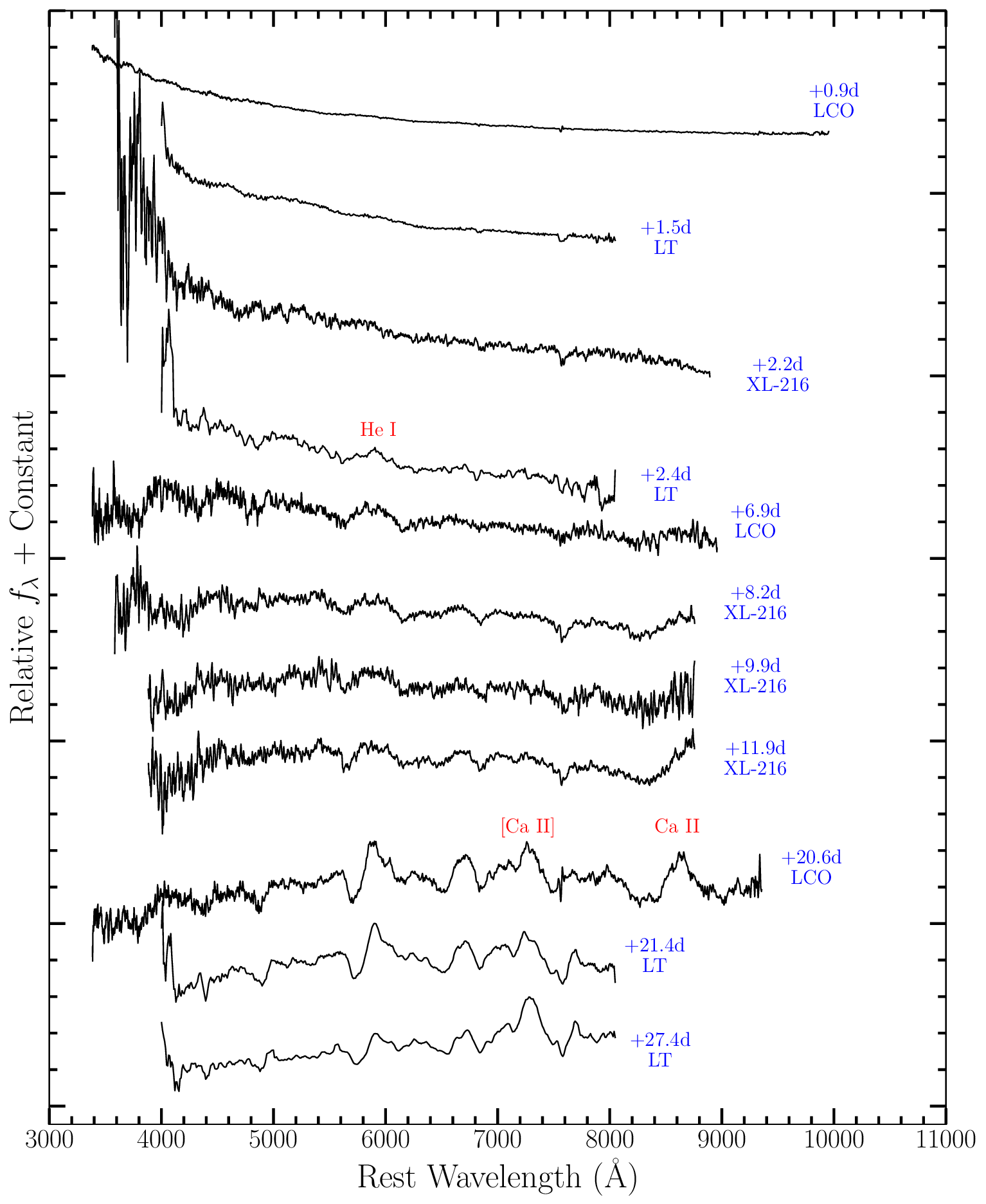}
\caption{Spectroscopic evolution of SN 2025coe with phases (blue) marked with respect to the first detection in the clear band during the photospheric phase.}
\label{spec_sum_photo}
\end{figure*}

\begin{figure*}[t!] 
\centering
\includegraphics[scale = 0.5]{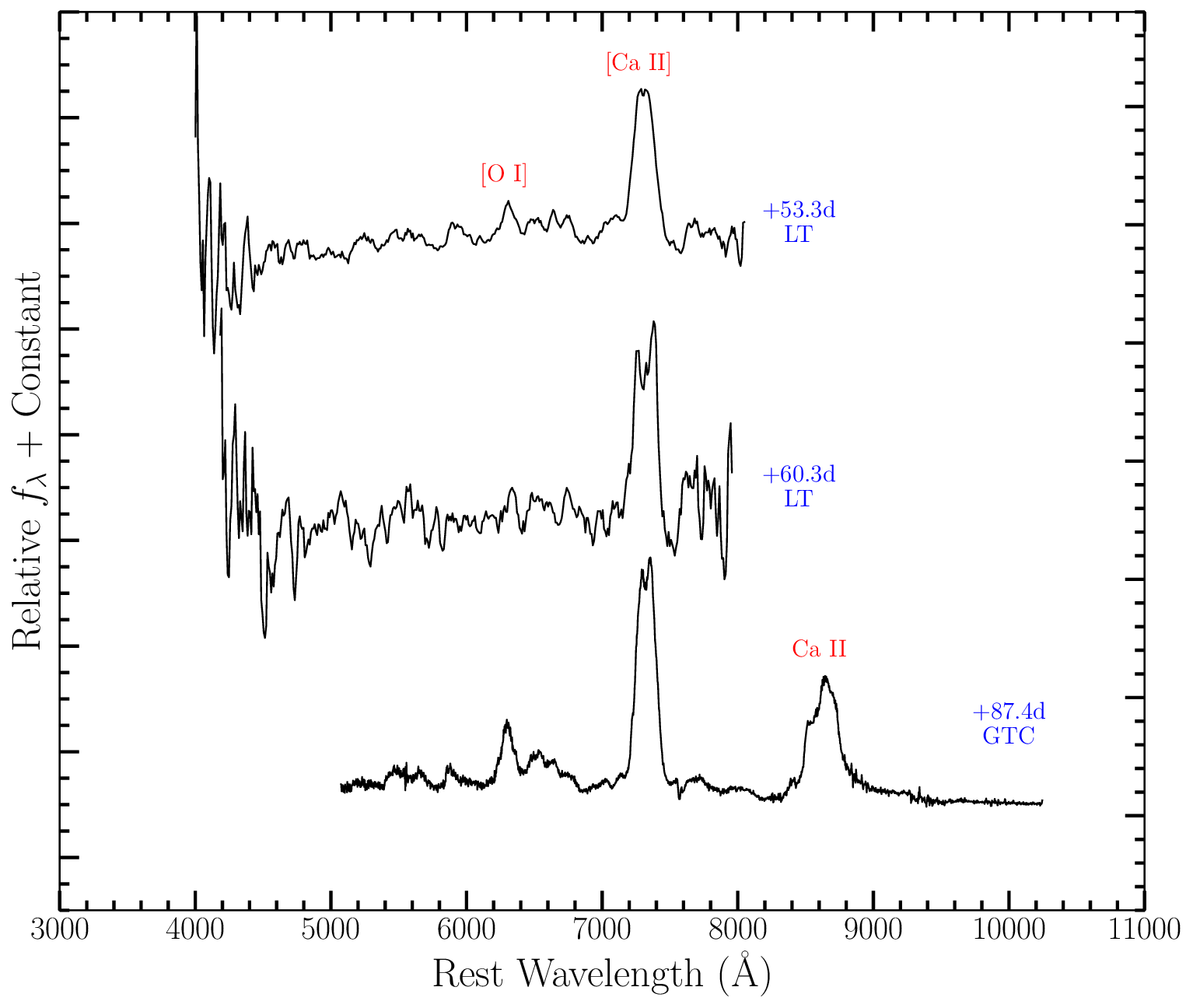}
\caption{Spectroscopic evolution of SN 2025coe with phases (blue) marked with respect to the first detection in the clear band during the nebular phase.}
\label{spec_sum_neb}
\end{figure*}

The spectroscopic evolution of SN 2025coe is presented in Figures \ref{spec_sum_photo} and \ref{spec_sum_neb}. The first three spectra exhibit a featureless continuum from $t$ = 0.9 days to $t = 2.2$ days. There are no significant narrow emission lines from the flash-ionization of the circumstellar medium (CSM) by the hot SN radiation or from ejecta-CSM interaction.

At t = 2.4 days, P-Cygni profiles emerge in the spectrum, specially in the He I $\lambda$5876 line. These features persist until $t = 27.4$ days. We measure the absorption component of the He I $\lambda$5876 P-Cygni profile using Gaussian profile fitting after removing the continuum baseline. The resulting line velocity $v_{\rm He \, I}$ is presented in Figure \ref{v_ph}. It shows a rapid decrease from $\sim$ 22,000 $\rm{km \, s^{-1}}$ at t = 2.4 days to $\sim$ 6,500 $\rm{km \, s^{-1}}$ by t = 10 days. Thereafter, $v_{\rm He \, I}$ remains nearly constant at $\sim$ 6,500 $\rm{km \, s^{-1}}$ until $t = 27.4$ days.

At the time of $t = 20.6$ days, the forbidden emission line of calcium, [Ca~II] $\lambda\lambda$7291, 7324, begins to emerge, when the ejecta are still optically thick given the P-cygni profiles of He I 5876. The P-cygni profile has disappeared at $t = 53.3$~days, signifying that the ejecta have transitioned to an optically thin state, marking the onset of the nebular phase in spectral evolution. During this phase, the spectrum exhibits strong [Ca II] lines and relatively weak [O I] $\lambda$$\lambda$6300, 6374 emission lines. These characteristics are fully consistent with the defining features of CaSTs: (1) rapid transition to the nebular phase ($\sim\!$\, a few $\times$ 10 days); (2) high flux ratio ${\text{[Ca~II]}}/{\text{[O~I]}}$ during the nebular phase.

\begin{figure}  [!t]
\centering
\includegraphics[scale = 0.3]{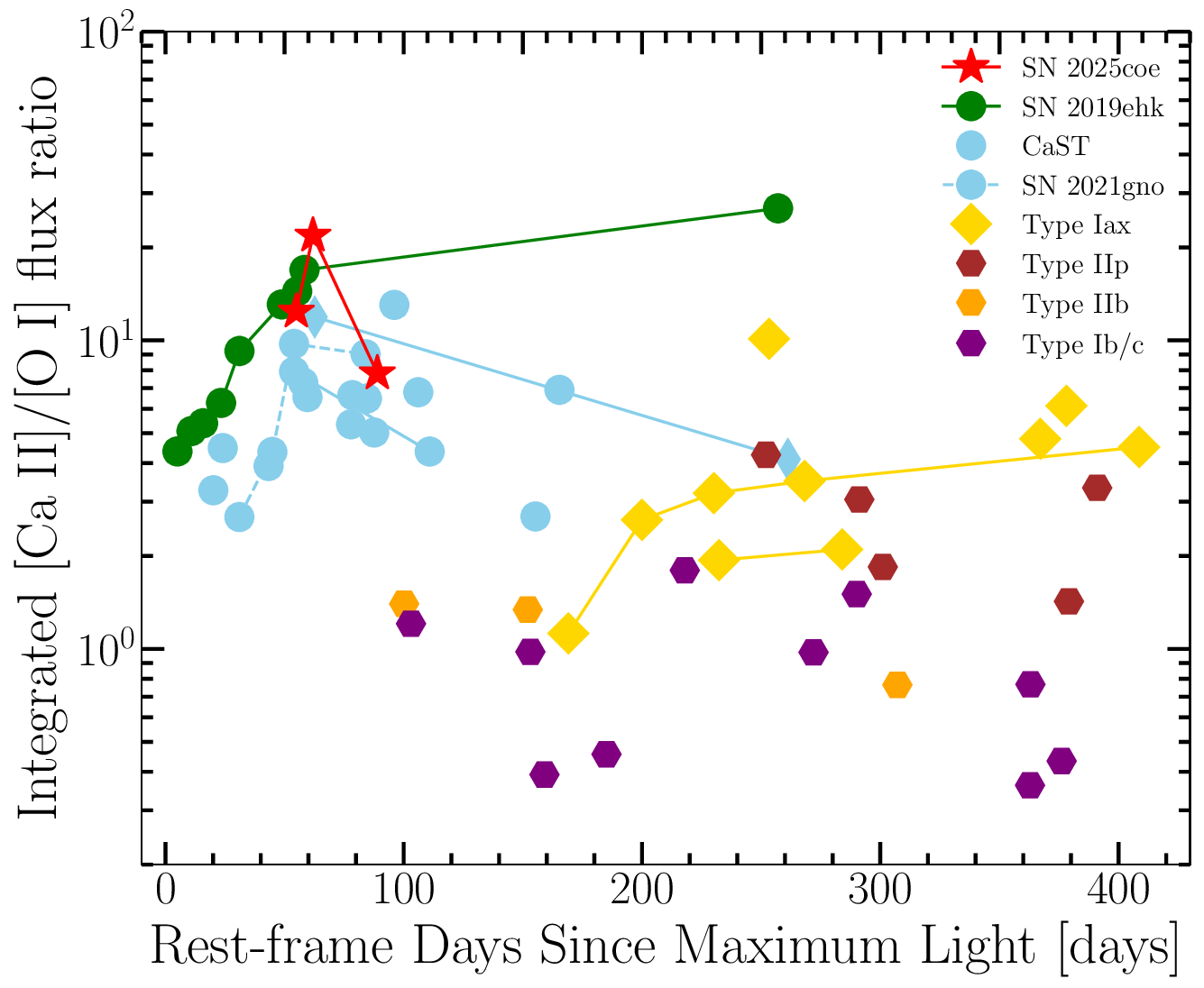}
\caption{The [Ca II]/[OI] flux ratio of SN 2025coe in comparison with other CaSTs and different SN subtypes.}
\label{Ca_O_ratio}
\end{figure}

Figure \ref{Ca_O_ratio} compares the [Ca II]/[O I] flux ratio of SN 2025coe with those of other CaSTs and SN subtypes. The [Ca II]/[O I] flux ratio of SN 2025coe exhibits an initial rise followed by a decline, resembling the behavior seen in another CaST, SN 2021gno \citep[]{Jacobson-Gal2022}. Moreover, at late phases, the [Ca II]/[O I] flux ratio of SN 2025coe declines rapidly, with a faster decay rate than any other CaSTs in the sample.

\section{Modeling the Bolometric Light Curve} \label{LC}



 

\begin{figure} [!t]
\centering
\includegraphics[scale = 0.4]{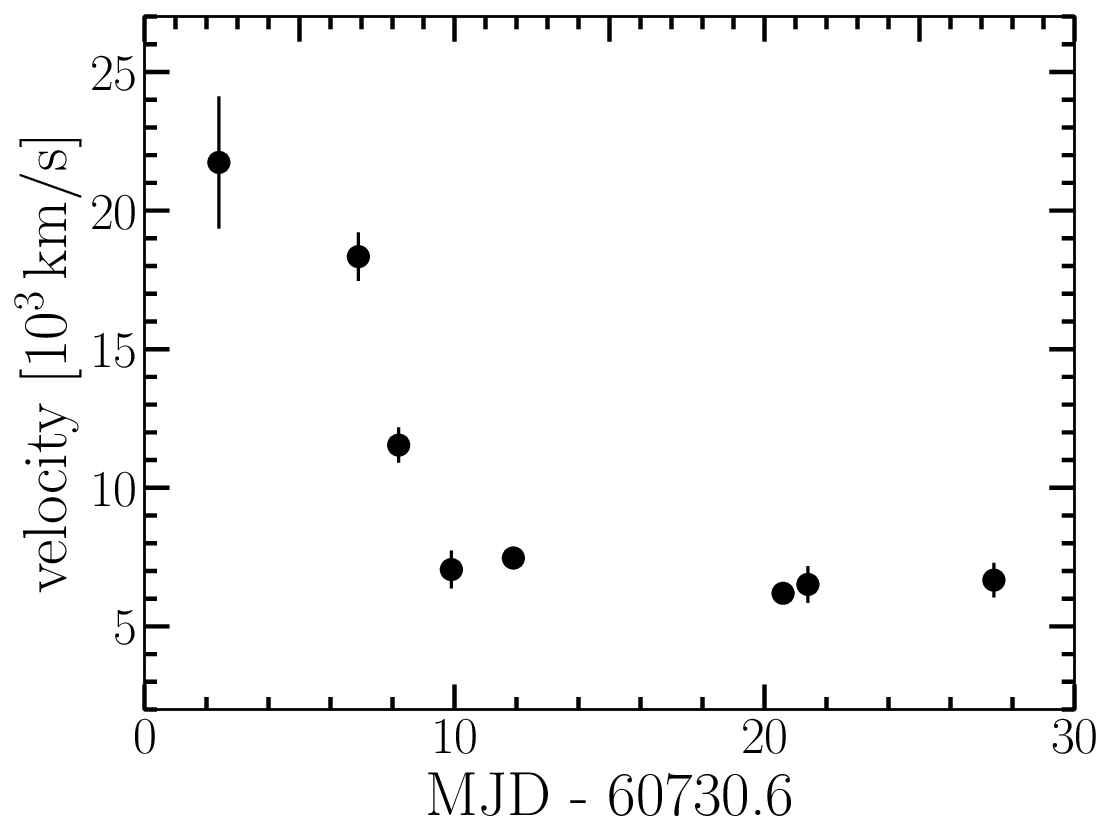}
\caption{Line velocities $v_{\rm He \, I}$ obtained from He I $\lambda$5876 absorption profile fitting.}
\label{v_ph}
\end{figure}

\begin{table} 
\centering 
\caption{Summary of Best-fitting Parameters:}
\begin{tabular}{cc} \hline 
\hline
parameters    & best-fit value  \\ 
\hline
$R_e \, (R_{\odot})$      & $13.5^{+64.1}_{-11.1} $  \\
$M_e \, (10^{-3} \, M_{\odot})$     &  $1.4^{+6.9}_{-1.2}   $ \\
$v_s \, (10^4 \, \rm km \, s^{-1})$     &   $0.8^{+3.8}_{-0.6} $ \\
$E_k \, (10^{50} \, \rm{erg})$   & $1.7^{+4.2}_{-1.5} $ \\
$M_{\rm ej} \, (M_{\odot})$    &  $0.29^{+0.14}_{-0.15} $ \\
$M_{\rm Ni} \, (10^{-2} \, M_{\odot})$   &  $2.4^{+0.06}_{-0.05} $ \\
\hline
\end{tabular}
\label{best-fit parameters}
\tablecomments{$R_e$: the radius of the extended material; $M_e$: the mass of the extended material; $v_s$: the shock velocity; $E_k$: the total kinetic energy of the ejecta, $M_{\rm Ni}$: the synthesized $^{56}$Ni mass; $M_{\rm ej}$: the total ejecta mass.}
\end{table}

We first focus on interpreting the physical origin of the first peak. Previous studies have suggested that the rapid post-peak decline observed in CaSTs may originate from either: (1) ejecta-circumstellar medium (CSM) interaction \cite[]{Jacobson-Gal2020}, or (2) shock breakout followed by rapid cooling \citep[]{De2018,Jacobson-Gal2020,Jacobson-Gal2022}. The absence of detectable narrow emission lines in our early-phase spectra provides no evidence for the presence of CSM.

In this work, we model the first bolometric light curve peak using the shock cooling model. When the high-velocity ejecta collide with the extended material surrounding the progenitor, the resulting shock rapidly heats the impacted material. The shock-heated material exhibits high initial velocities, as measured from He I absorption line fitting (Figure \ref{v_ph}). It then undergoes rapid adiabatic expansion and cooling, resulting in the observed steep decline in the early light curve. Here we adopt the shock cooling model described in \cite{Piro2015} to quantitatively analyze the early photometric evolution \citep[]{Padma2024}.

Besides, we interpret the main peak of SN 2025coe as being powered by the radioactive decay of $^{56}$Ni, consistent with other CaSTs, under the simplified assumption of the spherically symmetric ejecta. The light curve is analyzed separately for the early photospheric phase (t $<$ 30 days) using the analytic solutions from the Appendix A in \cite{Valenti2008} (based on \citey{Arnett1982}), and for the nebular phase (t $>$ 60 days) using the late-time solutions from the same reference (derived from \citey{Sutherland1984}, \citey{Cappellaro1997}, \citey{Colgate1997}).

We performed the combined model fitting to the bolometric light curve using the \texttt{emcee} Markov chain Monte Carlo (MCMC) sampler \citep[]{Foreman-Mackey2013}. Here, we adopt a constant opacity of $\kappa_{\rm sc} = 0.34 \, \rm{cm^2 \, g^{-1}}$ for the shock cooling model \cite{Piro2015}, and $\kappa_{\rm rd} = 0.1 \, \rm{cm^2 \, g^{-1}}$ in the radioactive decay model. The model constrains six physical parameters: the radius $R_e$ and the mass $M_e$ of the extended material, the shock velocity $v_s$, the total kinetic energy of the ejecta $E_k$, the synthesized $^{56}$Ni mass ($M_{\rm Ni}$), and the total ejecta mass ($M_{\rm ej}$). 

\begin{figure} [!t]
\centering
\includegraphics[scale = 0.35]{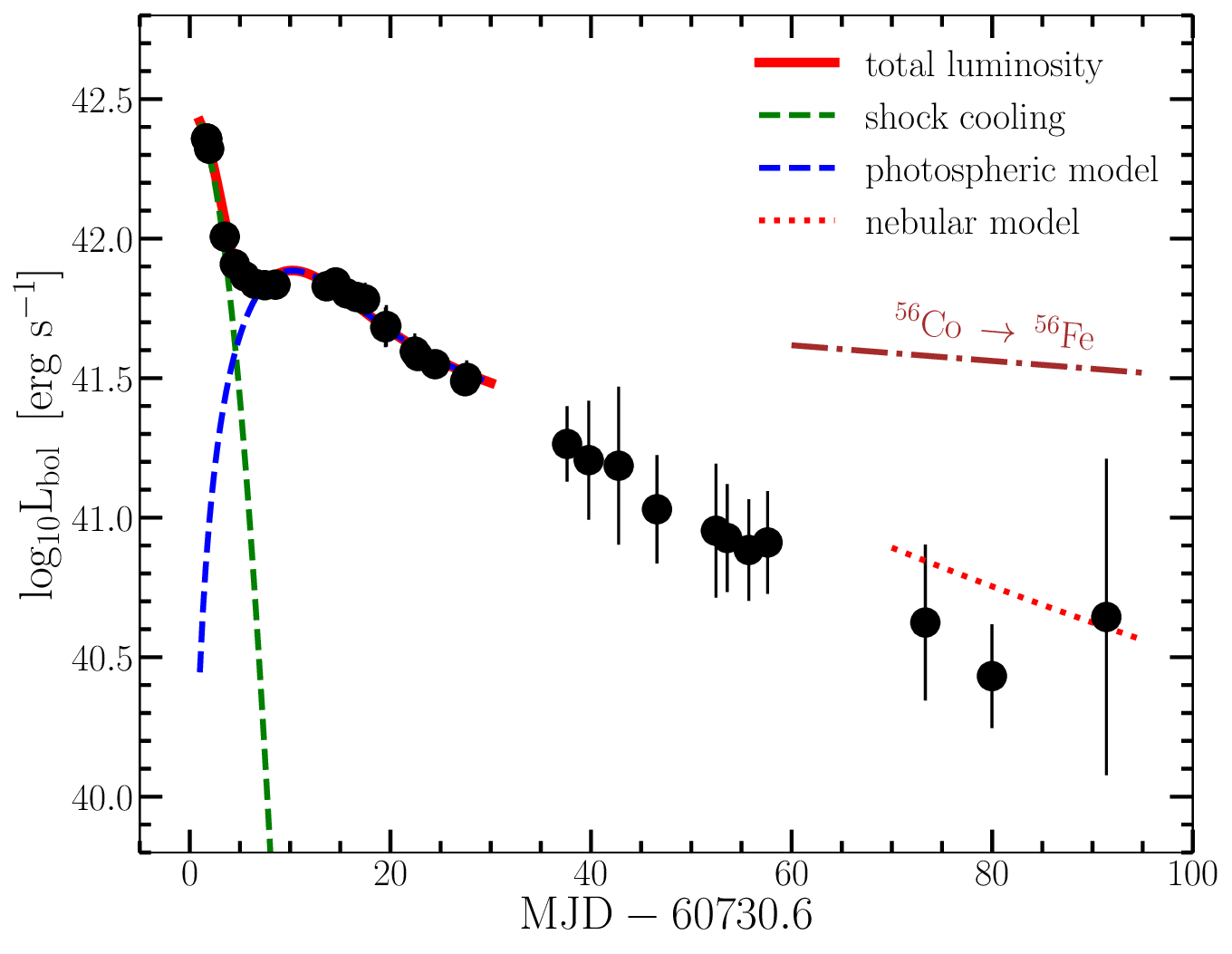}
\caption{The bolometric light curve of SN 2025coe is shown together with the best-fit model of shock cooling $+$ $^{56}$Ni radioactive decay, where the dot-dashed line represents the luminosity decline rate of the $^{56}$Co radioactive decay.}
\label{sc+Nipower}
\end{figure}

The best-fit results are presented in Figure \ref{sc+Nipower} and Table \ref{best-fit parameters}. The results reveal an extended material configuration with characteristic radius of $\sim 2.4 - 77.6 \, R_{\odot}$ and the mass of $\sim 0.2 - 8.3 \times 10^{-3} \, M_{\odot}$. The results also indicate a total ejecta mass of $\sim$ 0.29 $M_{\odot}$ with $^{56}$Ni mass of $\sim$ 0.02 $M_{\odot}$. The derived $^{56}$Ni-to-ejecta mass ratio of $\sim$ 0.08 is consistent with the low $^{56}$Ni fraction characteristic of the CaSTs population \citep[]{Perets2010, Alsabti2017}.

\section{Discussion}  \label{Discussion}
In this work, we present extensive photometric and spectroscopic monitoring of the CaST SN 2025coe. Our observations reveal a triple-peaked light curve and the first two peaks can be explained by a combination of shock cooling and radioactive heating. In this section, we discuss the progenitor properties of SN 2025coe and explore the physical origin of its possible third peak observed in multi-band light curves.

\subsection{Progenitor Properties}

SN 2025coe's remote location from the host galaxy and the host's old stellar population promptly exclude massive star explosion models \citep[]{Alsabti2017}. Table \ref{CaSTs_host} summarizes the host properties against other multi-peak CaSTs. A plausible progenitor model is the double WD merger model\cite[]{Bobrick2017,Perets2019,Zenati2019}. Recent simulations show that the complete tidal disruption of a hybrid HeCO WD by a low-mass CO WD can trigger a weak He detonation on the primary, producing slightly lower mass of the ejecta ($\sim \, 0.01-0.1 \, M_{\odot}$) and $^{56}$Ni yields $\sim$ a few $\times 10^{-4} - 10^{-3} \, M_{\odot}$. Such values are broadly consistent with our model results shown in Table \ref{best-fit parameters} \cite[]{Zenati2019}.

During mass transfer from the secondary to the primary prior to the disruption, the transferred material forms an accretion disk due to its sufficiently large angular momentum. \cite{Zenati2019} demonstrate that the inner region of the accretion disk may produce continuous outflow in the form of disk winds. The outflow expands outward to form CSM \citep[]{Raskin2013}. A fraction of the outflow might accumulate around the primary WD, forming a low-mass envelope \citep[]{Shen2012,Schwab2016}.

Our analysis of the bolometric luminosity of SN 2025coe reveals the presence of an extremely low-mass envelope surrounding its progenitor (see Table \ref{best-fit parameters}). The aforementioned double WD merger model provides a natural explanation for the formation of such an envelope. Notably, this model has also successfully accounted for the observed properties of other multi-peaked CaSTs \citep[]{Jacobson-Gal2020, Jacobson-Gal2022}.

\subsection{The {Potential} Third Peak}
We briefly examine the possible origin of the Potential third peak detected in the multi-band light curves of SN 2025coe. This feature is clearly visible in the $g$/Gbp/G/Grp/$I$ bands (see Figure \ref{phot_sum}), though our limited observational sampling in other wavelengths prevents further confirmation of its presence. The temporal coincidence between this peak and a weak bump in the photospheric temperature evolution (see middle panel in Figure \ref{BB_fit_res}) suggests a probably physical connection, likely involving the heating of the material.

A possible explanation involves interaction between the expanding ejecta and CSM or clump structures in the progenitor environment \citep[]{Jacobson-Gal2020}. The double WD merger model naturally provides a physical origin for the formation of such CSM/clump structures through pre-merger mass loss and tidal disruption processes. Such ejecta-CSM/clump interaction naturally accounts for both the observed photospheric temperature increase and the emergence of the third luminosity peak. {However, it is important to note that our spectroscopic sequence, presented in Figures \ref{spec_sum_photo} and \ref{spec_sum_neb}, lacks spectroscopic coverage at around 42.8 days. Furthermore, the spectra at $t = 27.3$ days and $t = 53.3$ days show no clear signs of ejecta-CSM/clump interaction. Therefore, }
this interpretation highlights the critical need for systematic, high-cadence late-time monitoring of CaSTs, which would enable more robust characterization of their progenitor systems through detailed studies of CSM/clump properties and mass-loss histories.

The detection of similar features in future CaST observations could provide important insights into the diversity of progenitor channels and their mass-loss mechanisms. In particular, coordinated multi-wavelength campaigns would help determine whether these third peaks [also detected in SN 2019ehk, \citep[]{Jacobson-Gal2020}] represent a common phenomenon among CaSTs or are specific to certain progenitor configurations.

\begin{table}[!t] 
\centering 
\caption{Summary of the host properties of multi-peaked CaSTs.}
\begin{tabular}{cccc} \hline 
\hline
Object name & Host galaxy & Host type & Projected offset \\ \hline
iPTF 16hgs\footnote{\cite{De2018}} & SDSS J005052 &  $-$  &   6 kpc \\
SN 2018lqo\footnote{\cite{De2020}} & CGCG 224-043 & E  & 15.5 kpc \\
SN 2019ehk\footnote{\cite{Jacobson-Gal2020}} & NGC 4231 & SAB(s)bc & 1.8 kpc  \\
SN 2021gno\footnote{\cite{Jacobson-Gal2022, Ertini2023}} & NGC 4165 & SAB(r)a  & 3.6 kpc  \\
SN 2021inl\footnote{\cite{Jacobson-Gal2022}} & NGC 4923 & E/S0   & 23.3 kpc \\
SN 2025coe & NGC 3277 & SAa/SAb & 39.3 kpc \\ \hline
\end{tabular}
\label{CaSTs_host}
\tablecomments{No systematic long-term photometric or spectroscopic follow-up observations are available for SN 2018lqo.}
\end{table}

\section{Conclusion} \label{Conclusion}
Our comprehensive study of SN~2025coe has revealed several remarkable properties that provide new insights into the nature of CaSTs. Through long-term photometric and spectroscopic observations, we have characterized this object as one of the members of its class. The key findings are summarized as follows:

\begin{itemize}
    \item
    SN~2025coe is located in the early-type galaxy NGC~3277 (morphological type SAa/SAb), with an exceptionally large projected physical offset of $\sim$ 39.3~kpc ($\sim 317.7$$"$) from the galactic center. This makes it the most distant multi-peaked CaST ever discovered relative to its host galaxy, suggesting its progenitor system might be ejected through dynamical interactions \citep[]{Alsabti2017}.

    \item
    Our multi-band monitoring campaign, spanning from $1.6$ to $\sim$ 90 days relative to the first detection, reveals that SN~2025coe reached a peak absolute magnitude of $M_{R, \, peak} \approx -15.8$~mag. The light curve exhibits a rapid post-peak decline rate of $\Delta m_{15}(R) \approx 1.5$~mag.

    \item
    The well-sampled multi-band light curves display {multiple} distinct peaks occurring at phases of  approximately $t \approx 1.6$ days, $t \approx 10.2$ days, {and a potential third peak at} $t \approx 42.8$~days relative to the first detection. This triple-peaked structure is exceptionally rare among known CaSTs, with only one other event (SN~2019ehk) showing similar features.

    \item
    Our spectroscopic sequence, beginning at just $+0.9$~days post-discovery, shows: (1) early-phase spectra dominated by broad (FWHM $\sim$22,000~km~s$^{-1}$) He I $\lambda$5876 emission; (2) the emergence of [Ca II] $\lambda\lambda$7291,7324 forbidden lines at $t = 20.6$~days while the spectrum still maintains photospheric characteristics of the He I P-Cygni profile; and (3) complete transition to the nebular phase by $t = 50.3$~days, marked by the disappearance of the He I P-Cygni profile.

    \item 
    The nebular-phase spectrum of SN 2025coe shows a [Ca II]/[O I] flux ratio of $> 2.0$, establishing SN~2025coe as a member of the CaST class. With its extensive photometric/spectroscopic sampling, it becomes the fifth confirmed case of CaSTs with at least two peaks, following iPTF 2018hgs, SN 2019ehk, SN 2021gno, and SN 2021inl. This growing sample enables statistical studies of their progenitor systems.

    \item
    Our bolometric light curve analysis using a combined shock-cooling and $^{56}$Ni-decay model yields best-fit parameters of $M_{\rm ej} \sim 0.29 M_{\odot}$ for the ejecta mass and $M_{^{56}\mathrm{Ni}} \sim 0.02 M_\odot$. The modeling further reveals the presence of a low-mass envelope ($M_e \approx 2.0 \times 10^{-4} - 8.3 \times 10^{-3} \, M_\odot$, $R_e \approx 2.4 - 77.6 \, R_\odot$) surrounding the progenitor. The CO WD + hybrid WD disruption scenario naturally explains: (1) the low ejecta and $^{56}$Ni masses, (2) the origin of the envelope, and (3) the formation of extended CSM/clump. We propose that ejecta-CSM/clump interaction may be responsible for producing the observed third peak in the multi-band light curves.
\end{itemize}

Future high-cadence transient surveys coupled with hydrodynamic simulations of such mergers will be crucial for testing whether this channel can account for the full diversity of observed CaST properties. {In particular, well-sampled monitoring from ultra-early to late phases—as exemplified by facilities like The Mini-SiTian Array \citep{Han2025}, which can provide earlier detection and higher-cadence observations—will be essential. This must be complemented by real-time, multi-band follow-up capable of dynamically increasing the cadence of photometric and spectroscopic observations in response to potential rebrightening or new peaks. Such a comprehensive approach is crucial for constraining the properties of the potential CSM/clump structures, thereby enabling a detailed reconstruction of the pre-explosion circumstellar environments of CaSTs.}

\section{ACKNOWLEDGEMENTS}
NCS is funded by the Strategic Priority Research Program of the Chinese Academy of Sciences Grant No. XDB0550300, the National Natural Science Foundation of China Grants No.12303051 and No. 12261141690, and the China Manned Space Program No. CMS-CSST-2025-A14. We acknowledge the support by the National Astronomical Research Institute of Thailand under the project number TRTToO$\_$2025001 and TRTC12B$\_$003. CC acknowledges the support of the China Scholarship Council (grant n. 202506380129). D.A. acknowledges financial support from the Spanish Ministry of Science and Innovation (MICINN) under the 2021 Ram\'{o}n y Cajal program MICINN RYC2021-032609. FP acknowledges support from the Spanish Ministerio de Ciencia, Innovaci\'{o}n y Universidades (MICINN) under grant numbers PID2022-141915NB-C21. ZG acknowledges the support supported from the China-Chile Joint Research Fund (CCJRF No.2301) and the Chinese Academy of Sciences South America Center for Astronomy (CASSACA) Key Research Project E52H540301. ZG is funded by ANID, Millennium Science Initiative, AIM23-001. 
WXL is supported by NSFC (12120101003 and 12373010), the National Key R$\&$D Program (2022YFA1602902 and 2023YFA1607804), and Strategic Priority Research Program of CAS (XDB0550100 and XDB0550000).
LZW is sponsored by the National Natural Science Foundation of China (NSFC) grants No. 12573050,  the Chinese Academy of Sciences South America Center for Astronomy (CASSACA) Key Research Project E52H540301, and  in part by the Chinese Academy of Sciences (CAS) through a grant to the CASSACA.
BCW acknowledges support from the National Key R$\&$D Program of China (grant Nos. 2023YFA1609700, 2023YFA1608304), the National Natural Science Foundation of China (NSFC; grant Nos. 12090040, 12090041, 12403022), and the Strategic Priority Research Program of the Chinese Academy of Sciences (grant Nos. XDB0550000, XDB0550100, XDB0550102). ZXN acknowledges support from the NSFC through grant No. 12303039 and No. 12261141690. 
This work is part of the Project RYC2021-032991-I, funded by MICIU/AEI/10.13039/501100011033, and the European Union “NextGenerationEU”/PRTR 

We thank the staff at all participating observatories, including the Swift/UVOT facility, the Xinglong 2.16 m telescope, the Liverpool Telescope, the Thai Robotic Telescope network, and the Gran Telescopio Canaria, for their support during our observing campaigns.

We acknowledge the use of data and services provided by ATLAS, ZTF, the Transient Name Server (TNS). We also acknowledge Moira Andrews and Joseph Farah for obtaining the LCO spectrum and making it publicly available via the TNS. We acknowledge the use of open-source software packages including astropy \citep[]{Astropy2013, Astropy2018, Astropy2022}, AutoPhot \citep[]{Brennan2022}, emcee \citep[]{Foreman-Mackey2013}, IRAF \citep[]{Tody1986}, and PypeIt \citep[]{Prochaska2020}.




\bibliography{sample631}{}
\bibliographystyle{aasjournal}

\end{CJK*}
\end{document}